\documentclass[11pt,twoside]{article}

\usepackage{asp2004}
\usepackage{epsf}
\usepackage{psfig}
\usepackage{lscape}

\begin{document}
\title{Kappa-Mechanism Excitation of Retrograde Mixed Modes in B-Type Stars}   
\author{Rich Townsend}   
\affil{Bartol Research Institute, University of Delaware, Newark, Delaware 19716, USA}    

\begin{abstract} 
The stability of retrograde mixed modes in rotating B-type stars is
investigated. It is found that these modes are susceptible to
$\kappa$-mechanism excitation, due to the iron opacity bump at $\log T
\approx 5.3$. The findings are discussed in the context of the
pulsation of SPB and Be stars.
\end{abstract}




\section{Introduction}

\begin{figure}[t!]
\plotone{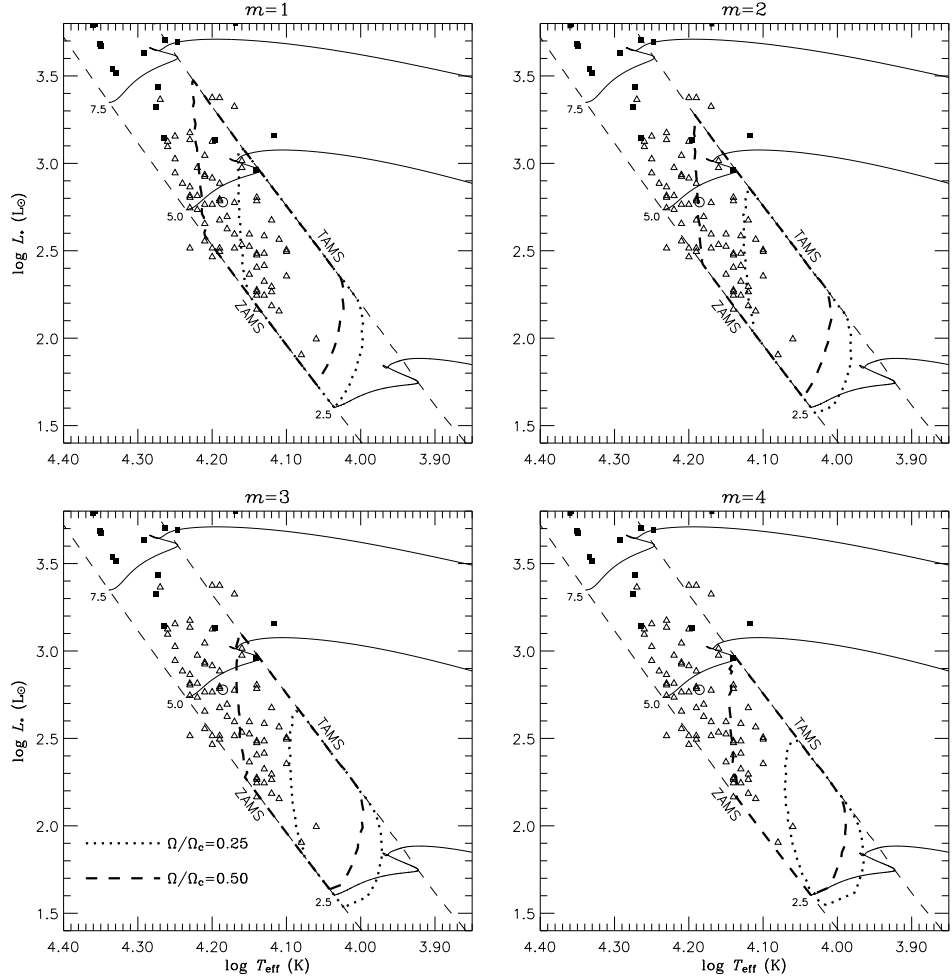}
\caption{Instability strips in the HR diagram for retrograde mixed
modes; the extent of the instability at the rotation rates
$\Omega/\Omega_{\rm c}=0.25$ (dotted) and 0.5 (dashed) is indicated
using thick lines. The thin dashed lines indicate the ZAMS and TAMS
boundaries, while the solid lines show tracks of three selected
evolutionary sequences, labeled at the ZAMS by their stellar mass. The
symbols indicate the inferred positions of selected SPB stars (open
triangles) and pulsating Be stars (filled squares).}
\label{fig:instab}
\end{figure}

In a uniformly-rotating star, the low-frequency oscillation spectrum
comprises four classes of pulsation mode --- Poincar\'{e}, Kelvin,
Rossby and mixed Rossby-Poincar\'{e}. The Poincar\'{e} modes rely on a
combination of buoyancy and the Coriolis force to restore displaced
fluid elements to their equilibrium positions; they reduce to ordinary
gravity (g) modes in the non-rotating limit. The Kelvin modes are a
prograde-propagating subset of the Poincar\'{e} modes that exhibit
geostrophic balance between polar pressure gradients and the polar
component of the Coriolis force. The Rossby modes, which take the form
of large-scale, retrograde-propagating horizontal circulations, arise
due to conservation of vorticity; they exist independent of buoyancy,
but reduce to trivial zero-frequency toroidal modes in the
non-rotating limit.

Finally, the mixed Rossby-Poincar\'{e} modes --- or `mixed modes' for
brevity --- are a curious hybrid between Poincar\'{e} and Rossby
modes. In the limit of slow rotation, the prograde mixed modes
(azimuthal orders $m < 0$) behave like Poincar\'{e} modes, while the
retrograde mixed modes ($m > 0$) behave like Rossby modes. However,
toward rapid rotation both prograde and retrograde mixed modes
approach one another in character, exhibiting properties that lie
somewhere between pure Rossby and Poincar\'{e} modes.

Like Rossby modes, non-trivial retrograde mixed modes do not exist in
the absence of stellar rotation. However, in contrast to the Rossby
modes (which are essentially solenoidal), these modes are able to
generate significant fluid compressions and rarefactions, due to the
Poincar\'{e}-mode character they acquire in the rapid-rotation
limit. This property renders them in principle susceptible to the
classical instability mechanisms ($\gamma$, $\delta$, $\epsilon$ and
$\kappa$) that rely on a thermodynamic Carnot cycle. The present paper
investigates whether such mixed-mode instability could be present in
B-type stars.

\section{Stability Analysis}

I use an updated version of the \textsc{boojum} pulsation code
\citep{Tow2005a} to investigate the stability of retrograde
$m=1\ldots4$ mixed modes in a set of stellar models representative of
main sequence B-type stars. This non-adiabatic code treats the effects
of the Coriolis force using the traditional approximation \citep[see,
e.g.,][]{LeeSai1987}, but departures from sphericity due to the
centrifugal force are neglected. Two rotation rates are considered:
$\Omega/\Omega_{\rm c} = 0.5$ (where $\Omega$ is the star's angular
velocity, and $\Omega_{\rm c}$ the corresponding critical angular
velocity) is chosen to approximate the observed upper limit on the
rotation of the slowly pulsating B-type (SPB) stars \citep{Mat2001},
while $\Omega/\Omega_{\rm c} = 0.25$ is representative of
more-moderately rotating stars.

The general finding is that the retrograde mixed modes are unstable
due to an opacity peak situated at a temperature $\log T \approx
5.3$. This peak is the same `iron bump' responsible for the pulsation
of $\beta$ Cep and SPB stars \citep[e.g.,][]{DziPam1993,Dzi1993}.  As
shown in Fig.~\ref{fig:instab}, the mixed-mode instability extends
over an effective temperature range $\log T_{\rm eff} \approx
4.0$--4.2, corresponding loosely to spectral types B4 to A0.

\section{Discussion}

Figure~\ref{fig:instab} indicates that the mixed-mode instability is
unlikely to furnish a better explanation for SPB-star variability than
the canonical picture of $\kappa$-mechanism excitation of
gravity/Poincar\'{e} modes. A similar conclusion could be reached for
the Be stars, were it not for the fact that these stars are rapid,
perhaps near-critical rotators \citep{Tow2004}. Although there are
difficulties in performing stability analyses at such extreme rotation
rates, the trends shown in Fig.~\ref{fig:instab} suggest that the
mixed-mode instability may shift to encompass the earlier spectral
types ---where the preponderance of Be stars are situated --- as
$\Omega$ approaches $\Omega_{\rm c}$. Therefore, it seems plausible
that the retrograde-propagating pulsation observed in variable Be
stars \citep{Riv2003} may actually be due to unstable mixed modes
rather than the usually-supposed g modes.

If confirmed, the latter hypothesis may help establish what role
pulsation plays in the Be phenomenon. Although they propagate with a
retrograde phase velocity, the unstable mixed modes show a prograde
group velocity, and therefore can assist in ejecting material into
orbit from the equator of a near-critical star
\citep[see][]{Owo2005}. This issue is discussed further in a
forthcoming paper \citep{Tow2005b}, which also presents a more
in-depth analysis of the newly-discovered mixed-mode
instability. Those interested in this topic should also refer to the
investigation by Saio (these proceedings), who examines retrograde
mixed modes (in his terminology, `buoyant r modes') using an approach
that does not rely on the traditional approximation.

\acknowledgements 

This research has been partially supported by US NSF grant AST-0097983
and NASA grant LTSA04-0000-0060.




\begin{thebibliography}{}
\bibitem[{{Dziembowski} {et~al.}(1993){Dziembowski}, {Moskalik}, \&
  {Pamyatnykh}}]{Dzi1993}
{Dziembowski}, W.~A., {Moskalik}, P., \&  {Pamyatnykh}, A.~A. 1993, \mnras,
  265, 588
\bibitem[{{Dziembowski} \& {Pamiatnykh}(1993)}]{DziPam1993}
{Dziembowski}, W.~A., \&  {Pamiatnykh}, A.~A. 1993, \mnras, 262, 204
\bibitem[{{Lee} \& {Saio}(1987)}]{LeeSai1987}
{Lee}, U., \&  {Saio}, H. 1987, \mnras, 224, 513
\bibitem[{{Mathias} {et~al.}(2001){Mathias}, {Aerts}, {Briquet}, {De Cat},
  {Cuypers}, {Van Winckel}, {Flanders.}, \& {Le Contel}}]{Mat2001}
{Mathias}, P., {Aerts}, C., {Briquet}, M., {De Cat}, P., {Cuypers}, J., {Van
  Winckel}, H., {Flanders.}, \&  {Le Contel}, J.~M. 2001, \aap, 379, 905
\bibitem[{{Owocki}(2005)}]{Owo2005}
{Owocki}, S. 2005, in ASP Conf. Ser. 337: The Nature and Evolution of Disks
  Around Hot Stars, {Ignace}, R., \&  {Gayley}, K.~G., eds., 101
\bibitem[{{Rivinius} {et~al.}(2003){Rivinius}, {Baade}, \& {{\v
  S}tefl}}]{Riv2003}
{Rivinius}, T., {Baade}, D., \&  {{\v S}tefl}, S. 2003, \aap, 411, 229
\bibitem[{{Townsend}(2005{\natexlab{a}})}]{Tow2005a}
{Townsend}, R.~H.~D. 2005{\natexlab{a}}, \mnras, 360, 465
\bibitem[{{Townsend}(2005{\natexlab{b}})}]{Tow2005b}
{Townsend}, R.~H.~D. 2005{\natexlab{b}}, \mnras, in press, doi
  10.1111/j.1365-2966.2005.09585.x
\bibitem[{{Townsend} {et~al.}(2004){Townsend}, {Owocki}, \&
  {Howarth}}]{Tow2004}
{Townsend}, R.~H.~D., {Owocki}, S.~P., \&  {Howarth}, I.~D. 2004, \mnras, 350,
  189
\end{thebibliography}
\end{document}